\documentclass[a4paper,11pt]{article}
\pdfoutput=1 

\usepackage{jheppub} 

\usepackage[T1]{fontenc} 

\def\be {\begin{equation}}
\def\ee {\end{equation}}
\def\bea {\begin{eqnarray}}
\def\eea {\end{eqnarray}}
\def\nn {\nonumber}

\definecolor{purple}{rgb}{0.7,0,1}

\title{Radiating black holes in Einstein-Maxwell-dilaton theory and cosmic censorship violation}

\author[a]{Pedro Aniceto,}
\author[b,a]{Paolo Pani}
\author[c]{and Jorge V. Rocha}

\affiliation[a]{CENTRA, Departamento de F\'{\i}sica, Instituto Superior T\'ecnico, Universidade de Lisboa,\\Avenida~Rovisco Pais 1, 1049 Lisboa, Portugal}
\affiliation[b]{Dipartimento di Fisica, ``Sapienza'' Universit\`a di Roma \& Sezione INFN Roma1,\\Piazzale Aldo Moro 5, 00185, Roma, Italy}
\affiliation[c]{Departament de F\'isica Fonamental, Institut de Ci\`encies del Cosmos (ICCUB), Universitat de Barcelona, Mart\'i i Franqu\`es 1, E-08028 Barcelona, Spain}

\emailAdd{pedro.aniceto@tecnico.ulisboa.pt}
\emailAdd{paolo.pani@roma1.infn.it}
\emailAdd{jvrocha@icc.ub.edu}

\abstract{
We construct exact, time-dependent, black hole solutions of Einstein-Maxwell-dilaton theory with arbitrary dilaton coupling, $a$. For $a=1$ this theory arises as the four-dimensional low-energy effective description of heterotic string theory. These solutions represent electrically charged, spherically symmetric black holes emitting or absorbing charged null fluids and generalize the Vaidya and Bonnor-Vaidya solutions of general relativity and of Einstein-Maxwell theory, respectively. The $a=1$ case stands out as special, in the sense that it is the only choice of the coupling that allows for a time-dependent dilaton field in this class of solutions. As a by-product, when $a=1$ we show that an electrically charged black hole in this theory can be overcharged by bombarding it with a stream of electrically charged null fluid, resulting in the formation of a naked singularity. This provides an example of cosmic censorship violation in an exact dynamical solution to low-energy effective string theory and in a case in which the total stress-energy tensor satisfies all energy conditions. When $a\neq1$, our solutions necessarily have a time-independent scalar field and consequently cannot be overcharged.
}

\begin{document} 
\maketitle
\flushbottom

\section{Introduction}\label{sec:intro}

One century has elapsed since Einstein first presented his theory of general relativity (GR) to the world~\cite{Einstein:1915ca}. Despite its overwhelming success, GR faces serious challenges at both the very large (cosmological) length scales and the very small (quantum) length scales. While in the former case the pressing issues of the nature of dark matter and dark energy may be conceivably resolved by postulating the existence of additional fields, for the latter the concern is the difficulty in combining GR with quantum mechanics and therefore obtain a complete theory of quantum gravity. Among the several proposals, string theory is one of the main contenders~\cite{Polchinski:2015pzt}. At any rate, any theory of quantum gravity certainly requires going beyond general relativity.

Models of gravity coupled to scalar fields and Maxwell terms are the simplest extensions of GR. Moreover, they arise naturally in supergravity theories that represent low-energy effective descriptions of string theories~\cite{Maharana:1992my,Sen:1994fa,Cvetic:1995kv,Ortin:2004ms}.
If string theory really is the theory of quantum gravity realised in nature, then classical physics operating in regions of weak gravity should be addressed in the context of such effective theories.
In particular, string effects on the gravitational collapse and black hole (BH) formation should be studied within this framework, at least for the sub-Planckian phase of the evolution.
These processes are fundamentally dynamic and exact, time-dependent solutions are typically hard to find\footnote{See however Ref.~\cite{Maeda:2005ci} for spherical collapse and black hole formation in Gauss-Bonnet gravity.}.

In this paper we partially remedy this situation by presenting time-dependent black hole solutions of Einstein-Maxwell-dilaton theory. For a certain value of the dilaton coupling constant, this theory is obtained as the four-dimensional low-energy effective description of heterotic [$E_8\times E_8$ or $SO(32)$] string theory, upon truncation of some form fields. However, several other values of the coupling are naturally generated by considering intersecting branes in string theory. In the interest of generality, we obtain solutions for arbitrary dilaton coupling. 
These novel spacetimes are the analogues of the well-known Vaidya~\cite{Vaidya:1951zz,Vaidya:1951zza,Lindquist:1965zz} and Bonnor-Vaidya~\cite{Bonnor:1970zz} solutions in Einstein gravity and Einstein-Maxwell theory, respectively.

Static, charged, spherically symmetric solutions of this theory have been known for some time~\cite{Gibbons:1987ps,Garfinkle:1990qj}. They differ from Reissner-Nordstr\"om black holes ---~their Einstein-Maxwell counterparts~--- in notable ways. For example, they possess only one horizon. Also, in the extremal limit they become singular and (for certain values of the dilaton coupling) their temperature becomes non-zero, or even infinite, though they are prevented from radiating by an equally infinite potential barrier~\cite{Holzhey:1991bx}.

We will restrict our study to spherically symmetric solutions, which are of interest for the simplest dynamical scenarios, namely spherical collapse. As we will see, the role of the ``radiation'' is played by a charged null fluid component that permeates spacetime, in full analogy with~\cite{Bonnor:1970zz}. Clearly, if we were in vacuum, i.e., if the stress-energy tensor vanished, there could be no gravitational or electromagnetic waves emitted since that would correspond to monopolar radiation, which is absent for massless spin-2 and spin-1 fields\footnote{Correspondingly, in spherical symmetry Birkhoff's theorem and Gauss' law force the metric and the Maxwell field to be static in vacuum.}. However, there can be scalar emission even when restricting to spherical symmetry, and indeed this is what one expects from, say, a dynamical spherical thin shell that is charged under the dilaton field~\cite{Cawley:1968}.

The solutions we present have a time-dependent metric and Maxwell field, but ---~for strictly positive dilaton couplings $a\neq1$~--- a time-independent dilaton\footnote{Note that the uniqueness theorems of Refs.~\cite{Mars:2001pz,Gibbons:2002ju} assume staticity, which is not the case in our study.}. The value $a=1$ is therefore special, in the sense that it is the only coupling which allows for time dependence in the scalar field, at least for the class of solutions we present. The other special value of the coupling constant is $a=0$, in which case the dilaton decouples and our solutions reduce to the Bonnor-Vaidya expressions. Hence, for $a\neq1$ these solutions cannot be used to describe a radiating scalar-charged shell collapsing in empty space. They can, however, be used to study the collapse of scalar-uncharged shells and thus serve as toy models for evaporating black holes in Einstein-Maxwell-dilaton theory, along the lines of Ref.~\cite{Farley:2005mw}. 

As we will address in section~\ref{sec:CCCtest}, these solutions can be employed in simple tests of the cosmic censorship conjecture (CCC), analogous to Ref.~\cite{Sullivan:1980} in Einstein-Maxwell theory. For the general case $a\neq1$ we prove that our black hole solutions cannot be overcharged by bombarding them with a spherically symmetric stream of charged null dust, satisfying standard energy conditions. Remarkably, for $a=1$ we show that there exists a family of solutions that describe initially regular states evolving into a final naked singularity. These solutions obey the usual energy conditions and thus constitute a violation of cosmic censorship within Einstein-Maxwell-dilaton theory.

We note that cosmic censorship in effective dilaton gravity was studied some years ago by Maeda {\it et al.}~\cite{Maeda:2001ie}, with a setup very similar to ours. However, there are two important differences: they considered asymptotically de Sitter solutions ---~and, as those authors showed, a non-vanishing cosmological constant does have a crucial effect on gravitational collapse~---, while we have a charged null fluid component which is absent in their setting.

For choices of the dilaton coupling parameter for which the Einstein-Maxwell-dilaton theory derives from a (consistent truncation of a) higher-dimensional string theory in the low energy limit, it is of obvious interest to know whether there exist, within the parent theory, natural sources capable of supporting our new solutions. I.e., are there objects in string theory whose four-dimensional low-energy effective description coincides with a charged null fluid? While some candidates present themselves, currently the answer is not obvious and a full analysis of this problem is beyond the scope of this paper. Nevertheless, we discuss this point at more length in the final section.

The rest of the paper is organised as follows. In the next section we display the field equations for the Einstein-Maxwell-dilaton theory. In section~\ref{sec:static} we briefly review the well known static black hole solutions of the theory under consideration. The derivation of the new time-dependent solutions is presented in section~\ref{sec:newsols}, together with their physical interpretation. In section~\ref{sec:CCCtest} we use these solutions to perform a simple test of the CCC in the context of low-energy effective string theory. We close in section~\ref{sec:conc} with a discussion of the results, their interpretation and a brief outlook.
We provide some details on the determination of energy conditions in the Appendix.

\section{Field equations}\label{sec:fieldeqs}

We consider the following Lagrangian for Einstein-Maxwell-dilaton theory (henceforth we set $G=c=1$),
\be
{\cal L} =\frac{\sqrt{-g}}{16\pi}\left[R-2(\nabla\phi)^2-e^{-2 a\phi}F^2+16\pi A_\mu J^\mu\right] + {\cal L}_{\rm m}\,,\label{lagrangian}
\ee
where $g$ is the determinant of the metric $g_{\mu\nu}$, $A_\mu$ is a Maxwell field with field strength $F_{\mu\nu}=\partial_\mu A_\nu-\partial_\nu A_\mu$, and $\phi$ is a scalar field (the dilaton), which is coupled to the field strength. For the sake of generality, we have parametrized the dilaton coupling by a constant $a$, with $a=1$ in the case of heterotic string theory and $a=\sqrt{3}$ in a Kaluza-Klein reduction from five dimensions~\cite{Gibbons:1987ps}. Besides these, other values can also be obtained from compactification (and truncation) of intersecting brane solutions~\cite{Papadopoulos:1996uq,Emparan:2004wy}. In the action above we have also included the minimal coupling of the Maxwell field to a current $J^\mu$ and an extra matter Lagrangian which will account for the fluid. The field equations arising from Eq.~\eqref{lagrangian} read 
\begin{subequations}\label{eq:fieldeqs}
\bea
&&\nabla_\mu \left( e^{-2 a \phi} F^{\mu \nu} \right) = -4\pi J^\nu\,,\label{MaxwellEOM}\\
&&\nabla^2 \phi + \frac{a}{2}e^{-2 a\phi} F_{\mu \nu} F^{\mu \nu} = 0\,,\label{dilatonEOM}\\
&&G_{\mu\nu} = 8\pi  T_{\mu\nu} \equiv 8\pi \left(T_{\mu\nu}^{\rm (dil)} + T_{\mu\nu}^{\rm (EM)} + T_{\mu\nu}^{\rm (fluid)}\right)\,,\label{EinsteinEOM}
\eea
\end{subequations}
where $G_{\mu\nu} := R_{\mu\nu} - \frac{1}{2}R g_{\mu\nu}$ is the Einstein tensor.
We split the total stress energy tensor $T_{\mu\nu}$ into three pieces according to their different origins: a contribution from the dilaton, $8\pi T_{\mu\nu}^{\rm (dil)} := 2\nabla_\mu\phi\nabla_\nu\phi - g_{\mu\nu}(\nabla\phi)^2$, a contribution from the electric field, $8\pi T_{\mu\nu}^{\rm (EM)} := e^{-2a\phi} \left( 2F_{\mu \alpha} F_\nu^\alpha -\frac{1}{2}g_{\mu \nu}F^2 \right)$, and the charged fluid energy-momentum tensor,
\be
T_{\mu\nu}^{\rm (fluid)}= T_{\mu\nu}^{\rm m} + g_{\mu\nu}A_\sigma J^\sigma - 2 A_{(\mu} J_{\nu)}\,,
\label{FluidTmunu}
\ee
where $T_{\mu\nu}^{\rm m}:= -\frac{2}{\sqrt{-g}} \frac{\partial {\cal L}_{\rm m}}{\partial g^{\mu\nu}}$.
We recover the Einstein-Maxwell theory by consistently setting the dilaton and its coupling to zero, $\phi=0=a$.

\section{Static black hole solutions}\label{sec:static}

Static solutions of the field equations~\eqref{eq:fieldeqs} in the absence of the source terms $J^\mu$ and $T_{\mu\nu}^{\rm m}$ were found in Refs.~\cite{Gibbons:1987ps,Garfinkle:1990qj}.
The electrically charged solution reads~\cite{Garfinkle:1990qj}
\begin{eqnarray}
 ds^2 &=& -\lambda dt^2 + \lambda^{-1} dr^2 + r^2 Bd\Omega^2 \,, \label{metricGHS} \\
 F &=&-\frac{Q}{r^2} dt\wedge dr\,,\label{FaradayGHS} \\
 e^{2\phi(r)}&=& e^{2\phi_0}(1-r_-/r)^\frac{2a}{1+a^2}\,, \label{dilatonTdependent} 
\end{eqnarray}
with
\begin{eqnarray}
 \lambda(r)&=& (1-r_+/r)(1-r_-/r)^\frac{1-a^2}{1+a^2}\,, \label{lambdaGHS}\\
 B(r)&=& (1-r_-/r)^\frac{2a^2}{1+a^2}\,, \label{BGHS}
\end{eqnarray}
where the physical mass $M$, the electric charge $Q$, and the dilatonic charge $D$ are related to $r_\pm$ by
\begin{equation}
 M=\frac{r_+}{2}+\frac{1-a^2}{1+a^2}\frac{r_-}{2}\,,\qquad Q^2=e^{2a\phi_0} \frac{r_+ r_-}{1+a^2}\,,\qquad D=\frac{a}{1+a^2}r_-\,. \label{MQD}
\end{equation}
For $a=0$ we simply recover the Reissner-Nordstr\"om solution, supplemented by a constant (decoupled) scalar field, with an event horizon at $r=r_+$ and an inner Cauchy horizon at $r=r_-$. But for any nonvanishing $a$ the surface $r=r_-$ is singular, since its area goes to zero. The absence of a naked singularity thus imposes $r_+>r_-$.

While the above solution~(\ref{metricGHS}-\ref{dilatonTdependent}) refers to electrically charged black holes, it is easy to obtain magnetically charged solutions via electric-magnetic duality~\cite{Garfinkle:1990qj}. It turns out that the metric remains unaltered while the dilaton field flips sign.

For any given choice of coupling constant $a$ ---~which distinguishes between different theories~--- these solutions are parametrized by two numbers, $r_\pm$\footnote{One may also count the asymptotic value of the dilaton, $\phi_0$, as a further parameter. However, this can be trivially generated since the sourceless field equations~\eqref{eq:fieldeqs} are invariant under a simultaneous shift of the dilaton, $\phi\to\phi+\phi_0$, and a rescaling of the Maxwell field, $A_\mu\to e^{a\phi_0}A_\mu$.}. Therefore, although there are three conserved charges ($M$, $Q$, $D$) they are not all independent. Indeed, they satisfy
\be
a^2 e^{-2a\phi_0} Q^2 = 2aMD - (1-a^2)D^2\,.
\label{eq:QMDconstraint}
\ee

In the following we will consider the generalization of these {\it static} solutions to time-dependent ones. In particular we will search for spherically symmetric solutions of the charged Vaidya type.

\section{Derivation of new time-dependent solutions}\label{sec:newsols}

We now adapt the procedure of Refs.~\cite{Bonnor:1970zz,Husain:1995bf} to the Einstein-Maxwell-dilaton theory we are considering.

Let us take the above static solution written in retarded/advanced Eddington-Finkelstein coordinate $u:= t-\epsilon r_*$ (with $dr/dr_*=\lambda$ and $\epsilon=\pm1$ for retarded and advanced coordinate, respectively), and promote the mass and electric charge to functions of $u$, so that $r_+=r_+(u)$ and $r_-=r_-(u)$ in the equations above. 
In practice, our ansatz is
\bea
 ds^2 &=&-\lambda(u,r)du^2-2\epsilon dudr+r^2 B(u,r)d\Omega^2\,,   \label{ansatzEF}\\
 F &=& -\frac{Q(u)}{r^2} du\wedge dr\,,
\eea
where $\lambda(u,r)$, $B(u,r)$, $Q(u)$ [and similarly $\phi(u,r)$] are obtained from the expressions in Eqs.~\eqref{dilatonTdependent}--\eqref{MQD} by promoting $r_+$ and $r_-$ to functions of $u$. Following the static case, our solutions will only be parametrized by two functions of $u$. Nevertheless, one might expect that the most general such solution is parametrized by three functions.

With the ansatz above, Maxwell's equations impose 
\be
 J^\nu = - \frac{e^{-2a\phi(u,r)}}{4\pi r^2}Q'(u) \delta^\nu_r\,, \label{Jmu}
\ee
where the prime denotes a derivative with respect to $u$.
Thus, as long as the electric charge is not constant we will get a non vanishing radial current, which is nevertheless divergence-free, $\nabla_\nu J^\nu=0$. 
The electromagnetic field is Coulombian and the electric current decays as $r^{-2}$ at large $r$. 

On the other hand, the dilaton field equation~\eqref{dilatonEOM} imposes
\be
a \left(1-a^2\right) r_-'(u) = 0\,.
\ee
Therefore, if $a=\pm1$ or $a=0$ the dilaton equation is automatically satisfied for any choice of $r_-(u)$, whereas for $a^2\neq0,1$ it enforces $r_-(u)={\rm constant}$\footnote{Henceforth we restrict to $a\geq0$. This condition can be enforced without loss of generality through the symmetry $\phi\to-\phi$, $a\to-a$ of the Lagrangian~\eqref{lagrangian}.}. Leaving aside the Einstein-Maxwell limit $a=0$, which was analyzed in Ref.~\cite{Bonnor:1970zz}, the case of heterotic string theory, $a=1$, is therefore special and will be treated separately.

Now we turn our attention to the Einstein field equations~\eqref{EinsteinEOM}.
It is useful to introduce the following (future-pointing in contravariant form) null vectors:
\be
\ell_\mu = -\partial_\mu u = -\delta_\mu^u\,, \qquad 
n_\mu= \frac{1}{2} g_{uu} \delta_\mu^u - \epsilon \delta_\mu^r \,,
\ee
which satisfy $\ell_\mu\ell^\mu = 0$, $n_\mu n^\mu=0$ and $\ell_\mu n^\mu=-1$.
We seek a solution of the Einstein equations with an {\it additional} source term (i.e., besides the stress-energy tensor due to the dilaton and the Maxwell field) of the form of a null fluid (cf. e.g.~\cite{Husain:1995bf}):
\be \label{NullFluid}
8\pi T_{\mu\nu}^{\rm (fluid)} = \mu \ell_\mu\ell_\nu + (\rho+P) (\ell_\mu n_\nu + \ell_\nu n_\mu) + P g_{\mu\nu}\,,
\ee
where $\mu$ receives contributions from both the matter Lagrangian, ${\cal L}_{\rm m}$, and from the current terms, $A_{(\mu} J_{\nu)}$, in Eq.~\eqref{FluidTmunu}. Note that, for our solutions, the term $A_\sigma J^\sigma$ in Eq.~\eqref{FluidTmunu} vanishes and the term $A_{(\mu} J_{\nu)}$ only has a $uu$-component, which contributes to the energy density $\mu$ but not to $P$ and $\rho$.

The above form~\eqref{NullFluid} of the stress-energy tensor is a generalization of a null dust ---~which would have $P=\rho=0$ and thus describes a pressureless fluid with energy density $\mu$ moving with four-velocity $\ell^\mu$. Nevertheless, this stress-energy tensor supports an energy flux only along the null vector $n$, namely
\be
T_{\mu\nu}^{\rm (fluid)} \ell^\mu\ell^\nu = 0\,, \qquad
8\pi T_{\mu\nu}^{\rm (fluid)} n^\mu n^\nu = \mu\,.
\ee

The energy conditions for such a stress-energy tensor have been discussed in Refs.~\cite{Husain:1995bf, Wang:1998qx}. If $\mu\neq0$ the stress-energy tensor is of type~II according to the standard terminology of Refs.~\cite{Hawking:1973uf, Kuchar:1990vy}. In this case the dominant energy condition imposes $\mu>0$ and $\rho\geq P\geq0$, whereas the weak and strong energy conditions both impose $\mu>0$, $\rho\geq 0$ and $P\geq0$.

\subsection{Solving the Einstein field equations with a null fluid}\label{sec:EnergyConds}

From Eq.~\eqref{EinsteinEOM}, it turns out that the total stress-energy tensor sourcing the Einstein equations,
\begin{equation}
T_{\mu\nu} = T_{\mu\nu}^{\rm (dil)}+T_{\mu\nu}^{\rm (EM)}+T_{\mu\nu}^{\rm (fluid)}\,,
\label{totalTmunu}
\end{equation}
cannot be written in the form~\eqref{NullFluid} unless $a=0$, in which case the dilaton component of the stress-energy tensor vanishes and we recover the Bonnor-Vaidya solutions~\cite{Bonnor:1970zz}. In this case the total stress-energy tensor can actually be written in the form~\eqref{NullFluid}, as shown in~\cite{Husain:1995bf}.

{For generic coupling $a$, both the electromagnetic component $T_{\mu\nu}^{\rm ({\rm EM})}$ and the fluid component $T_{\mu\nu}^{\rm (fluid)}$ are of the form~\eqref{NullFluid}. The stress-energy tensor derived from the Maxwell field has $\mu^{\rm (EM)}=0$ and $P^{\rm (EM)}=\rho^{\rm (EM)}$, whereas the fluid component has vanishing pressure and
\bea
\mu_{a=1}&=&\frac{\epsilon\left[(r_+(u)r_-(u))'-2 r r_+'(u)\right]+2r^2 r_-''(u)}{2r^2 [r-r_-(u)]} \,,\label{mua1} \\
\mu_{a=0}&=& -\epsilon\frac{r_+'(u)(r-r_-(u))+r_-'(u)(r-r_+(u))}{r^3} \,,\label{mua0} \\
\mu_{a\neq1,0}&=&-\epsilon\frac{\left[(1+a^2)r-r_-\right]}{(1+a^2)r^{\frac{3+a^2}{1+a^2}} \left[r-r_-\right]^{\frac{2a^2}{1+a^2}}} r_+'(u) \,,\label{mu}\\
\rho_{a=1}&=& -\epsilon\frac{1}{r[r-r_-(u)]} r_-'(u)\,, \label{rhoa1} \\
\rho_{a\neq1}&=&0\,. \label{rho}
\eea
When the dilaton coupling vanishes the equations above reproduce the Bonnor-Vaidya expressions, as expected.

For this type of stress-energy tensor {\it all} energy conditions reduce to $\mu\geq0$ and $\rho\geq0$. The latter condition is automatically satisfied when $a\neq1$, whereas when $a=1$ it imposes
\be
\epsilon r_-'(u) \leq 0 \qquad \textrm{for} \;\; a=1\,, \label{EC1}
\ee
since the denominator in Eq.~\eqref{rhoa1} is necessarily positive for the metric to be regular. Likewise, when $a\neq1$ Eq.~\eqref{mu} imposes 
\be
 \epsilon r_+'(u)\leq0 \qquad \textrm{for} \;\; a\neq1\,, \label{EC2b}
\ee
i.e., the (past) apparent horizon must increase (decrease) in time for absorbing (radiating) solutions, for which $\epsilon=-1$ ($\epsilon=+1$), as expected.
For $a=1$ we obtain instead, from Eq.~\eqref{mua1},
\be
2r^2 r_-'' + \epsilon \left[ (r_+ r_-)' - 2r r_+' \right] \geq 0 \qquad \textrm{for} \;\; a=1\,. \label{EC2c}
\ee

The relations between $r_\pm$ and $(M, D)$ allow us to express Eq.~\eqref{mua1} alternatively as
\be
\mu_{a=1}=\frac{2}{r^2 \left(r-2D\right)} \bar{\mu}\,,
\ee
where we defined
\be
\bar{\mu} := r^2 D'' + \epsilon MD' - \epsilon (r-D)M' \,.
\ee
Therefore, in the special case $a=1$, the energy conditions can be reduced to
\be
\epsilon D' \leq0\,, \qquad  \bar{\mu}\geq0\,. \label{EC2}
\ee
We remark, in passing, that for $a=1$ the existence of radiating ($\epsilon=+1$) solutions for which the mass {\em grows} in time is allowed in principle ---~at least momentarily~--- as long as the second derivative of the dilaton charge is sufficiently large (and positive). This curious observation is similar to the Einstein-Maxwell case, where the Schwarzschild mass can also increase if the black hole ionizes at a faster rate~\cite{Bonnor:1970zz}.

In the analysis above we took a practical approach, considering the functions $r_\pm(u)$ are known {\it a priori} and deriving from them the stress-energy tensor components. Conversely, given a choice of matter content ($\mu$ and $\rho$) and the parameters of an initially static BH ($r_\pm$) one can integrate Eq.~\eqref{rhoa1}, for $a=1$, to obtain the function $r_-(u)$. Upon inserting the result into Eq.~\eqref{mua1}, one can integrate to determine $r_+(u)$ and therefore the entire evolution of the system. Likewise, for $a\neq1$ and $a\neq0$, Eq.~\eqref{mu} can be integrated for any given $\mu$ to determine $r_+(u)$ (note that $r_-={\rm const}$ in this case). We remark that this discussion does not apply to the case $a=0$.

\subsection{The constant-$D$ solution}\label{sec:constant-D}

As previously discussed, the case $a=1$ is special since it is the only non-trivial coupling which allows for time dependence of the scalar charge $D(u)=r_-(u)/2$. However, also in this case a particularly simple solution is obtained when $D={\rm const}$, i.e. $r_-'(u)=0$. For such a choice the dilaton field~\eqref{dilatonTdependent} becomes time-independent and the sole time dependence is in the metric and in the Maxwell field. 

In this situation the $a=1$ and $a\neq1$ cases can be treated simultaneously and Eq.~\eqref{mu} and \eqref{rho} are valid for any $a$. In particular, $\rho=P=0$ for any $a$ and so the fluid stress-energy tensor is actually of the form of {\it null dust}, just like in the Vaidya~\cite{Vaidya:1951zz,Vaidya:1951zza} and Bonnor-Vaidya~\cite{Bonnor:1970zz} solutions.

The condition~\eqref{EC1} is trivially satisfied and the second condition~\eqref{EC2b} only imposes that the (past) apparent horizon shrinks in time for the radiating ($\epsilon=1$) solution and increases in time for the absorbing ($\epsilon=-1$) solution.

It is illustrative to show the constant-$D$ solution for $a=1$ explicitly, to which we restrict for the rest of this subsection. In such case $M(u)=r_+(u)/2$, $Q^2(u)/M(u)=2 e^{2\phi_0} D ={\rm const},$\footnote{For generic $a\neq1$ the constant-$D$ condition can be equivalently written as $M(u)-\sqrt{M(u)^2-(1-a^2)e^{-2a\phi_0}Q(u)^2}= \textrm{const}$.} and
\begin{flalign}
& ds^2 = -\left( 1- \frac{2M(u)}{r} \right)du^2 - 2\epsilon dudr + r^2\left(1- \frac{2D}{r}\right) d\Omega^2\,,\label{metricTdependent2a1}\\
& F = -\frac{Q(u)}{r^2} du\wedge dr\,, \qquad e^{2\phi}= e^{2\phi_0}\left(1- \frac{2D}{r}\right)\,, \label{dilatonTdependent2a1}\\
&\mu(u,r) = -\epsilon \frac{2}{r^2}\frac{(r-D)}{r-2D}M'(u) \,, \label{mu_radiatinga1}
\end{flalign}
with $\rho=P=0$ and $J^\mu$ given by Eq.~\eqref{Jmu} with $a=1$. Finally, the condition $\mu\geq0$ reduces to $\epsilon M'(u)\leq0$, i.e. for $\epsilon=1$ (respectively, $\epsilon=-1$) the mass decreases (respectively, increases) in time, as one would expect for a radiating (respectively, absorbing) solution.

In the previous subsection we discussed the energy conditions on the fluid's stress-energy tensor. However, if the solution as a whole is to make physical sense then the {\em total} stress-energy tensor ---~including contributions from the charged null dust, from the Maxwell field and from the dilaton~--- should satisfy the usual energy conditions. Comfortingly, this is guaranteed if the energy conditions on the null dust component are obeyed. It is well known that the electromagnetic field complies with the dominant and the weak energy conditions ---~although they can be easily violated by scalar fields~\cite{Hawking:1973uf,Barcelo:2000zf}~--- and it can be checked explicitly that the weak, strong and dominant energy conditions on the total stress-energy tensor are satisfied. We refer to the Appendix for details on the energy conditions.

\subsection{Physical interpretation of the constant-$D$ solution}\label{sec:interpretation}

It turns out that the three contributions to the stress-energy tensor~\eqref{totalTmunu} are not individually conserved:
\begin{subequations}
\begin{flalign}
& 8\pi\chi \nabla^\mu T_{\mu\nu}^{\rm (dil)} =  \left\{0,\frac{2 r_- r_+(u)}{r(r-r_-)},0,0\right\}\,, \\
& 8\pi\chi \nabla^\mu T_{\mu\nu}^{\rm (EM)} =  -\left\{\frac{1+a^2}{a^2} r_+'(u),\frac{2 r_- r_+(u)}{r(r-r_-)},0,0\right\}\,, \\
& 8\pi\chi \nabla^\mu T_{\mu\nu}^{\rm (fluid)} =  \left\{\frac{1+a^2}{a^2} r_+'(u) ,0,0,0\right\}\,,
\end{flalign}
\end{subequations}
where $\chi:=  \left(1-r_-/r\right)^{\frac{2 a^2}{a^2+1}}[\left(1+a^2\right)^2 r^4]/(a^2 r_-)$.
However, their sum is conserved, of course, by virtue to the contracted Bianchi identity.

Recall this solution is supported by null dust with energy density $\mu$ propagating with four-velocity $\ell^\mu$. Therefore, the total flux ${\cal F}$ of energy across a sphere of constant $r$ is given by
\begin{flalign}
{\cal F} &= \text{vol}(S^2)\, (-{{T^r}_u}^{\rm (fluid)}) = 4\pi r^2 B(r) \epsilon \frac{\mu(u,r)}{8\pi} \nn\\
&=-\frac{(1+a^2)r-r_-}{2(1+a^2)r}r_+'(u) = - \frac{d}{du} \left[ M - \frac{Q^2 e^{-2a\phi_0}}{2r} \right]\,,
\label{flux}
\end{flalign}
where in the last step we used the relations~\eqref{MQD} between $(M,Q)$ and $(r_+,r_-)$ and the fact that $r_-={\rm const}$ for the constant-$D$ solution. Note there is no contribution to the flux of energy from the Maxwell field, ${{T^r}_u}^{\rm (EM)}=0$. There is also no contribution from the dilaton, since it is static. Therefore, this is really the {\it total} energy flux.

The expression on the right hand side of Eq.~\eqref{flux} is none other than the variation of the {\it energy contained inside} the sphere of radius $r$. From the total energy in the spacetime, $M$, one must subtract the electric energy stored outside the sphere. To support our assertion, we then have to compute the energy density in the electromagnetic field, which is given by
\be
8\pi T^{\rm (EM)}_{\mu\nu} \xi^\mu \xi^\nu = \rho^{\rm (EM)} = e^{-2a\phi(r)}\left(\frac{\partial A_u}{\partial r}\right)^2\,,
\ee
where $\xi^\mu = \left(-g_{uu}\right)^{-1/2} \delta_u^\mu$ is the normalized four-vector of an observer sitting at constant $r$ coordinate, in addition to constant angular coordinates. The electric energy stored in the space outside a sphere of radius $r$ is obtained by integration over the volume,
\bea
E^{\rm (EM)}_{>r} &=& 4\pi \int_r^{\infty} T^{\rm (EM)}_{\mu\nu} \xi^\mu \xi^\nu \; \bar{r}^2 B(\bar r) d\bar r  = \frac{e^{-2a\phi_0} Q^2}{2r} \,, \label{Esphere}
\eea
which precisely matches the second term inside the brackets in Eq.~\eqref{flux}.

The energy stored inside a sphere of a given radius $r$ 
is a coordinate-dependent quantity, whereas the total energy in the spacetime---which is obtained by taking the $r\to\infty$ limit---has a clear invariant meaning.
The latter agrees with the result obtained by using any standard pseudotensor method~\cite{Chamorro:1994vw}. However, we note that when considering strictly {\em finite regions} of spacetime (even for asymptotically flat spacetimes) different pseudotensor methods typically yield different order $O(r^{-1})$ terms for the energy and power radiated~\cite{Virbhadra:1991ap,Radinschi:2006xm} and generally they will not agree with the subleading term in Eq.~\eqref{flux}.

We can also establish a parallel with the physical interpretation given in~\cite{Sullivan:1980}, applied to the absorbing ($\epsilon=-1$) constant-$D$ solution.
The second term inside the brackets in the right-hand side of Eq.~\eqref{flux} is related to the work, $dW$, done on the charge $dQ$ (assumed to have the same sign of $Q$ so that $Q^2$ increases) by the electrostatic repulsive force, as the spherical distribution of charges move from $\infty$ to a radius $r$. To see this, one first computes the Lorentz force:
\be
f_{(el)}^\alpha = g^{\alpha\mu} F_{\mu\nu}J^\nu = - J_u \frac{\partial A_u}{\partial r} \delta^\alpha_r = \frac{r_-}{8\pi (1+a^2) r^4}\left(1-\frac{r_-}{r}\right)^\frac{-2a^2}{1+a^2}r_+'(u) \delta^\alpha_r\,,
\ee
where we note that in this case $u$ is an advanced time coordinate. Then, by integrating over the whole sphere, the work done against the electrostatic force reads
\bea
W &=& \int_\infty^r f_{(el)}^r 4\pi {\bar r}^2 B(\bar r) d{\bar r} 
= -\frac{r_-}{2(1+a^2) r} r_+'(u) = -\frac{d}{du}\left(\frac{e^{-2a\phi_0} Q^2(u)}{2r}\right)\,,
\eea
where again in the last step we used the relations~\eqref{MQD} between $(M,Q)$ and $(r_+,r_-)$ in the particular case of the constant-$D$ solution.
This result matches the contribution of the electric field to the power radiated computed in Eq.~\eqref{flux}, which explains why the charged dust particles are not decelerated by the electric field and actually move along null geodesics.

Finally, let us compute the flux of charge across a sphere with radius $r$:
\be
{\cal J} = \text{vol}(S^2)\, J^r = -4\pi r^2 \left(1-\frac{r_-}{r} \right)^{\frac{2a^2}{1+a^2}} \frac{e^{-2a\phi} Q'}{4\pi r^2} = -e^{-2a\phi_0} Q' \,,
\ee
where Eq.~\eqref{dilatonTdependent} was used in the last step.
This is independent of $r$, which is a consequence of charge conservation.

\bigskip

To summarize, this solution is very similar to the Bonnor-Vaidya solution. As in the latter case, there is a null dust that carries away energy from the central body and a null electric current which removes charge (for radiating solutions, $\epsilon=+1$). The dilaton does not play any role because it is time-independent---and the mass and charge are linked through constraint~\eqref{eq:QMDconstraint}. Therefore, when $M$ decreases, $Q^2$ must also decrease.

\subsection{The non-constant-$D$ solution}\label{sec:nonconstant-D}

As previously discussed, the non-constant-$D$ solutions require $a=1$. The full solution reads
\begin{subequations}
\bea
ds^2 &=& -\left(1- \frac{2M(u)}{r} \right)du^2 + 2dudr + r^2\left(1- \frac{2D(u)}{r}\right) d\Omega^2\,, \label{metricTdependent2}\\
F&=& \frac{Q(u)}{r^2} du\wedge dr\,, \qquad\qquad
Q^2(u)= 2D(u)M(u) e^{2\phi_0}\,, \label{MaxwellTdependent2}\\
e^{2\phi}&=& e^{2\phi_0}\left(1- \frac{2D(u)}{r}\right)\,, \label{dilatonTdependent2}\\
J^\mu&=& -\frac{[D(u)M(u)]'}{4\pi r[r-2D(u)]Q(u)}\delta^\mu_r \,,\\
\mu&=&\frac{2\left\{ r^2 D''(u) +\epsilon M(u)D'(u) -\epsilon [r-D(u)]M'(u) \right\}}{r^2[r-2D(u)]} \,, \label{mu_absorbing}\\
\rho&=&\frac{-2\epsilon D'(u)}{r[r-2D(u)]}\,, \label{rho_absorbing}
\eea
\end{subequations}
and the fluid has vanishing pressure, $P=0$. The energy conditions for the fluid component of the stress-energy tensor impose that expressions~\eqref{mu_absorbing} and~\eqref{rho_absorbing} be non-negative.

In the Appendix we show that the energy conditions for the \emph{total} stress-energy tensor are satisfied throughout the entire spacetime provided the fluid satisfies the energy conditions. In other words, $\mu\geq0$ and $\rho\geq0$ in Eqs.~\eqref{mu_absorbing} and~\eqref{rho_absorbing} are also sufficient conditions for the regularity of the total matter content of the non-constant-$D$ solutions. It is worth pointing out that the dilaton component does violate energy conditions sufficiently close to the apparent horizon, as demonstrated in the Appendix. This is somewhat reminiscent of the situation for dilaton black holes in higher curvature gravity~\cite{Kanti:1995vq}, although in that case it is the total effective stress-energy tensor that violates energy conditions close to the horizon due to contributions of higher derivative terms.

\section{Violation of the cosmic censorship conjecture}\label{sec:CCCtest}

The absorbing solutions obtained in section~\ref{sec:newsols}, with $\epsilon=-1$, describe electrically charged BHs in Einstein-Maxwell-dilaton gravity that are bombarded with a spherically symmetric stream of charged null fluid.

These solutions are useful to perform tests of the CCC which, roughly speaking, asserts that curvature singularities developing during the evolution of regular initial data should be covered by event horizons in any physically realistic spacetime with sensible matter content~\cite{Penrose:1969pc}. In principle, if a black hole solution admits an extremal configuration, one can imagine an experiment in which the black hole is made to exceed extremality, e.g., by overcharging or overspinning it~\cite{Wald:1974}. With our exact solutions, one may attempt to violate the condition $r_+(u)>r_-(u)$ by investigating how the black hole reacts to the stream of charged null fluid (cf. e.g.~\cite{Sullivan:1980} for a similar gedankenexperiment in Einstein-Maxwell theory).

Let us start by considering the simplest case: the absorbing constant-$D$ solution for dilaton coupling $a=1$, which was presented explicitly in Eqs.~\eqref{metricTdependent2a1}--\eqref{mu_radiatinga1}, with $\epsilon=-1$. In this case the charged null fluid sourcing the solution reduces to charged null dust.

As previously discussed, all energy conditions impose $\mu>0$ in an open domain that contains the apparent horizon, $r_+=2M(u)$. Because $D$ is constant, if we start with a regular solution at $u=0$, i.e., $M(0)>D$, then $\mu(0,r\geq r_+)>0$ and Eq.~\eqref{mu_radiatinga1} implies $M'(0)>0$. This, in turn, imposes $M(u>0)>D$ in the future.  This simple argument shows that in this case the energy conditions enforce cosmic censorship.

It is also easy to generalize this argument to any value of $a$. In the generic case the energy conditions only impose Eq.~\eqref{EC2b} which, for $\epsilon=-1$, implies $r_+'(u)\geq0$. Since $r_-={\rm const}$, if we start with a regular solution, $r_+>r_-$, the energy conditions again prevent the formation of a naked singularity in the future.

Now we move on to the analysis of the non-constant-$D$ solutions discussed in Sec.~\ref{sec:nonconstant-D}. This case is less trivial and --- as we will see --- much more interesting, as it allows the formation of a naked singularity from a regular initial BH geometry.

Consider a solution with constant mass $M>0$, with non-decreasing scalar charge, $D'(u)\geq0$, and obeying
\be
4D''(u)\geq MD'(u)/D(u)^2\,.
\label{inequality4D}
\ee
One may take $D(u<0)={\rm const}$, so that the fluid is absent before $u=0$ and the spacetime is static at early advanced times $u<0$. We also require that $M>D(u<0)$ for the curvature singularity at $r=2D(u)$ to be covered by the horizon at $r=2M$. The energy condition arising from Eq.~\eqref{rho_absorbing} is automatically satisfied. Moreover, the energy condition stemming from Eq.~\eqref{mu_absorbing}, $\mu\geq0$, is also satisfied in the entire spacetime\footnote{Recall the radial coordinate is bounded from below by $2D(u)$, where spacetime ends.} by virtue of inequality~\eqref{inequality4D}. An example is provided by the following explicit solution,
\be
M={\rm const}\,, \qquad\qquad
D(u)=\left\{
  \begin{array}{ll}
    \frac{M}{2} & {\rm for} \; u<0\,,\\ 
    \left(1+\frac{u^2}{u_s^2}\right)\frac{M}{2} \quad & {\rm for}\; 0\leq u\leq u_s\,,
  \end{array}
\right.
\ee
which starts out as a regular black hole for $u<0$. The above solution satisfies Eq.~\eqref{inequality4D} for any $M\geq 3\sqrt{3}u_s/16$.
For $0\leq u<u_s$ the body accretes scalar charge (without increasing mass) supplied by an infalling charged null fluid satisfying all standard energy conditions. Finally, at $u=u_s$ the extremal value is attained, $D(u_s)=M$, and thus a naked singularity is formed.

We therefore reach the remarkable conclusion that cosmic censorship can be violated in Einstein-Maxwell-dilaton theory, for the particular dilaton coupling $a=1$ (which is arguably the most interesting case from a phenomenological point of view). The counterexamples to the CCC we discussed above require spacetimes with non-constant dilaton charge $D$, which do not appear in our class of solutions if $a\neq1$.
We stress that the energy conditions for the total stress-energy tensor are satisfied throughout the entire spacetime, whereas the dilaton component of $T_{\mu\nu}$ violates energy conditions sufficiently close to the apparent horizon (cf. the Appendix).

In the case in which we do find CCC violation, the singularity --- initially hidden inside the black hole --- is ``pushed outwards'' by virtue of the charged matter accreted, until it reaches the apparent horizon and becomes visible. Thus, the nature of the naked singularity obtained is distinct from that of shell-crossing singularities~\cite{Szekeres:1995gy, Joshi:2012ak}. We also point out that this outcome does not require any fine tuning of parameters. However, we have only considered spherically symmetric solutions and in this sense the violation is not guaranteed to be generic.

This situation may be compared with other tests of the CCC that use test bodies to attempt to overcharge or overspin black holes~\cite{Wald:1974,Hubeny:1998ga,Jacobson:2009kt,BouhmadiLopez:2010vc}. The main difference is that our analysis employs exact solutions of the field equations, whereas the latter studies adopt the test particle approximation, and therefore neglect finite-size and backreaction effects. Other non-perturbative tests of the CCC have been performed using thin shells~\cite{Boulware:1973,Hubeny:1998ga,Gao:2008jy} and taking static spacetimes to describe both the interior and exterior regions. Such approaches therefore assume no radiation is present.

Of course, even if cosmic censorship is violated in the low-energy effective Einstein-Maxwell-dilaton theory, this does not imply a violation of the CCC in string theory, because near the hypothetical naked singularity both curvature and string coupling become large~\cite{Horne:1993sy}. Therefore, $\alpha'$ (stringy) corrections and loop (quantum) corrections must be taken into account and might completely modify the outcome of gravitational collapse.

\section{Discussion}\label{sec:conc}

We have presented a family of time-dependent black hole solutions to a class of Einstein-Maxwell-dilaton theories with arbitrary dilaton coupling. The solutions are spherically symmetric and asymptotically flat. They can be either radiating or absorbing (depending on the choice of $\epsilon=\pm1$) and are determined by two free functions of the retarded/advanced time coordinate, $r_\pm(u)$. Alternatively, they are characterised by the total mass $M(u)$ and dilatonic charge $D(u)$.

The constant-$D$ solution is a physically sensible solution to the Einstein-Maxwell-dilaton field equations, which is also time-dependent, although the dilaton only has $r$-dependence ---~the dilaton charge is, therefore, constant. For $\epsilon=+1$, it becomes a radiating solution ($M'(u)\leq0$), with the energy (and charge) loss being accounted for by a null dust component in the stress-energy tensor.

For the particular case of coupling constant $a=1$ ---~and only in that case~--- the dilaton may be time-dependent. For example, we can have outgoing ($\epsilon=+1$) null fluid solutions with constant electric charge, $Q={\rm const}$. However, for these spacetimes the energy conditions impose $M'(u)\geq0$ so their interpretation as radiating solutions may not be appropriate.
More interestingly, we have shown that a subset of our absorbing solutions (with $\epsilon=-1$) describe an initially regular black hole that evolves into a naked singularity upon the accretion of a charged null fluid. The matter content supporting these solutions can be made to obey the standard energy conditions. Therefore, this constitutes a counterexample to the CCC in the context of Einstein-Maxwell-dilaton theory.

The question of whether the charged null dust sourcing the time-dependence of our solutions can be embedded in low-energy effective string theory after dimensional reduction remains open. A gas of 6-branes in heterotic string theory wrapped on a six-torus would, at first, appear to be a natural candidate but these objects are known not to be supersymmetric~\cite{Horowitz:1991cd} and therefore they would not behave as dust. Moreover, D-branes typically have non vanishing tension and are therefore not light-like objects\footnote{We thank R. Emparan for pointing this out.}. But at least null D-branes arise in string theory in the strong coupling limit~\cite{Lindstrom:1997uj,Bergshoeff:1998ha}.

Another reasonable possibility would be for the higher-dimensional Yang-Mills field to act as a source for our solutions. The Einstein-Maxwell-dilaton theory (for $a=1$) is obtained as the four-dimensional low-energy effective description of heterotic string theory, where the Maxwell field arises from a $U(1)$ subgroup of the (consistently truncated) non-abelian gauge group. If the remaining ten-dimensional gauge fields are not truncated they will naturally source the Maxwell field\footnote{We thank D. Mateos for suggesting this possibility.}. This clearly deserves further study.

Finally, we observe that charged null dust radiation can emerge in certain Kaluza-Klein reductions of higher-dimensional purely gravitational theories~\cite{Maeda:2006hj}, although this particular origin seems somewhat unnatural since the fundamental theory is Gauss-Bonnet gravity in six (or more) dimensions and several conditions must be imposed on the dimensional reduction scheme: for example, the compactification manifold should be a space of constant negative curvature satisfying a so-called Einstein-space condition, and the Gauss-Bonnet coupling constant can only take a specific value.

The constant-$D$ solution might also be useful to study the dynamics of a spherical thin shell around a static black hole in Einstein-Maxwell dilaton theory. Indeed, one may use such a solution to match with the static black hole interior as to allow a spherical collapsing shell to radiate. As discussed in the introduction, this would not be the expected description of a collapsing shell, charged under the dilaton field, since in that situation there would be scalar radiation and no gravitational or electromagnetic waves. Nevertheless, this can serve as a toy model to further address cosmic censorship in the vein of string theory. This problem is currently under study and will be reported elsewhere.

The solutions we presented are electrically charged. The magnetically charged version can be obtained by starting with the static magnetic black holes and will require the presence of (magnetically) charged null dust.

Another interesting application is to extend our results to asymptotically anti-de Sitter spacetimes. In the context of the AdS/CFT correspondence, Vaidya-like solutions have been investigated recently to study thermalization and entanglement in strongly coupled field theories (see Refs.~\cite{Bhattacharyya:2009uu,AbajoArrastia:2010yt,Keranen:2011xs,Liu:2013qca,Alishahiha:2014cwa,Fonda:2014ula} for an illustrative sample). Some of these models naturally contain a dilaton with a nontrivial self-potential and possible couplings to gauge fields, thus resembling closely the setup studied in the present work.

It would also be desirable to go beyond our solutions and obtain radiating/absorbing spacetimes with dynamics also in the dilaton field for $a\neq1$. More generically, are there dynamical solutions not constrained by Eq.~\eqref{eq:QMDconstraint}? Can the CCC be violated also in these cases? These are challenging questions that will probably require a numerical integration of the field equations.

\acknowledgments

We are indebted to Roberto Emparan and David Mateos for useful discussions and for comments on a draft version of this paper.
We also thank an anonymous referee for constructive comments, which prompted us to extend the analyis conducted in Sec.~\ref{sec:CCCtest}.
This work was supported by the European Community through
the Intra-European Marie Curie Contract No.~AstroGRAphy-2013-623439 and by FCT-Portugal through the project IF/00293/2013.
JVR acknowledges financial support from the European Union's Horizon 2020 research and innovation programme under the Marie Sk\l{}odowska-Curie grant agreement No REGMat-2014-656882.
JVR was partially supported by FPA2013-46570-C2-2-P.

\appendix
\section{Energy conditions on the total stress-energy tensor}

Here we present some details concerning the energy conditions on the total stress-energy tensor~\eqref{totalTmunu}, for the most interesting value of the dilaton coupling, $a=1$. For completeness, we also include results for the dilaton component of the stress-energy tensor in the end. The following analysis is {\em not} restricted to the constant dilaton case.

In general relativity the various energy conditions (weak, strong or dominant) can be expressed in terms of the eigenvectors and eigenvalues of the stress-energy tensor~\cite{Hawking:1973uf,Kuchar:1990vy}. This can be conveniently done by projecting onto an orthonormal basis. In our specific case there is one timelike eigenvector $E^{(0)}$ and three spacelike eigenvectors $E^{(1)}$, $E^{(2)}$ and $E^{(3)}$, which identifies the stress-energy tensor as being type I. Explicitly, the eigenvectors are given by
\bea
E^{(0)}_\mu &=& \frac{r_- \sqrt{r}}{\Theta^{1/4}} \left[\frac{r_-^2 (r-r_+)+\sqrt{\Theta}}{2 \epsilon\, r\, r_-^2}\delta^u_\mu +\delta^r_\mu\right] \,, \\
E^{(1)}_\mu &=& \frac{r_- \sqrt{r}}{\Theta^{1/4}} \left[\frac{r_-^2 (r-r_+)-\sqrt{\Theta}}{2 \epsilon\, r\, r_-^2}\delta^u_\mu +\delta^r_\mu\right] \,, \\
E^{(2)}_\mu &=& \sqrt{r(r-r_-)} \delta^\theta_\mu \,,\\
E^{(3)}_\mu &=& \sin(\theta)\sqrt{r(r-r_-)} \delta^\varphi_\mu \,,
\eea
where we have defined
\be \label{chi_u_total}
\Theta =r_-^2 \left[\left(2 \epsilon \, r^2  r_-'+r_- (r-r_+)\right)^2 + 4 r^2 (r-r_-) \left(2 r^2 r_-''-2\epsilon \, r\, r_+'+\epsilon ( r_+ r_-)'\right) \right] \,.
\ee
From Eq.~\eqref{mua1}, we note that the function $\Theta(u)$ is guaranteed to be non-negative if the fluid component is required to satisfy energy conditions, $\mu\geq0,\, \rho\geq0$.

Written in the basis $\{E^{(i)}, \; i=0,\dots,3\}$ the total stress-energy tensor is diagonal and the associated eigenvalues $\lambda_i$ are
\bea
\lambda_0 &=&     -\frac{2 r_- r_+(r-r_-) -4 \epsilon r^2 (r-r_-) r_-'+\sqrt{\Theta}}{4 r^3 (r-r_-)^2}     \label{EC_lambda0}       \,, \\
\lambda_1 &=&    -\frac{2 r_- r_+ (r-r_-) -4 \epsilon r^2 (r-r_-) r_-'  - \sqrt{\Theta}}{4 r^3 (r-r_-)^2} \,,\label{EC_lambda1} \\
\lambda_2 &=& \lambda_3 =  -\frac{r_- \left[2 r ( \epsilon r  r_-'-r_+)+r_- (r+r_+)\right]}{4 r^3 (r-r_-)^2}  \,.\label{EC_lambda2}
\eea
For such a type I tensor, $-\lambda_0$ corresponds to the proper energy density of the spacetime, while the $\lambda_i$ with $i=1,2,3$ yield the principal stresses.

We can now formulate the energy conditions for the total stress-energy tensor $T_{\mu\nu}$. The weak energy condition requires $-\lambda_0 \geq 0$ and $\lambda_i \geq \lambda_0$ for $i=1,2,3$. These inequalities are automatically satisfied by virtue of Eq.~\eqref{EC1} if we require the fluid component to satisfy the corresponding energy conditions.

In addition, the dominant energy condition imposes $-\lambda_0 \geq \lambda_i$ for $i=1,2,3$.
The strong energy condition requires instead $\sum_{i=1}^3 \lambda_i\geq\lambda_0$.
With a little effort it can be shown that all these inequalities are once again obeyed if Eq.~\eqref{EC1} is assumed to hold.

We conclude that the fulfilment of the energy conditions by the fluid stress-energy tensor are sufficient to guarantee that the total stress-energy tensor obeys them as well, at least for the Einstein-Maxwell-dilaton theory with coupling $a=1$.

\subsection{Energy conditions for the dilaton component of the stress-energy tensor}
Turning to the stress-energy tensor derived from the dilaton alone, $T_{\mu\nu}^{\rm (dil)}$, one finds that it is also of type I. Proceeding as above we find find that its eigenvalues are given by
\be
\lambda_0^{\rm (dil)} = -\lambda_1^{\rm (dil)} = \lambda_2^{\rm (dil)} = \lambda_3^{\rm (dil)} =
  -\frac{r_- \left(2 \epsilon\, r^2 r_-' + r_- (r-r_+)\right)}{4 r^3 (r-r_-)^2}  \,.
\ee
Therefore, the weak, dominant and strong energy conditions are satisfied provided
\be
2\epsilon\, r^2 r_-'+r_- \left(r-r_+\right)\geq 0 \,.
\ee
However, these conditions are manifestly violated sufficiently close to the apparent horizon $r=r_+$ if inequality~\eqref{EC1} is imposed, i.e., if the energy conditions for the null dust component are satisfied.

\bibliographystyle{JHEP}
\bibliography{bib}

\end{document}